\def\ls{\lower4pt\hbox{${\buildrel < \over \sim}$}}
\def\gs{\lower4pt\hbox{${\buildrel > \over \sim}$}}

\documentclass[12pt, preprint]{aastex}

\slugcomment{Accepted for publication in {\it The Astrophysical Journal}}

\shorttitle{3C~66A in 2007 -- 2008}
\shortauthors{B\"ottcher et al.}

\begin{document}

\title{The WEBT Campaign on the Intermediate BL Lac Object 
3C~66A in 2007 -- 2008\footnote{The radio-to-optical data 
presented in this paper are stored in the WEBT archive; for 
questions regarding their availability,  please contact the WEBT 
President Massimo Villata ({\tt villata@oato.inaf.it}).}}

\author{M. B\"ottcher\altaffilmark{1}, 
K. Fultz\altaffilmark{1}, 
H. D. Aller\altaffilmark{2}, 
M. F. Aller\altaffilmark{2}, 
J. Apodaca\altaffilmark{3},
A. A. Arkharov\altaffilmark{4}, 
U. Bach\altaffilmark{5},
R. Bachev\altaffilmark{6},
A. Berdyugin\altaffilmark{7},
C. Buemi\altaffilmark{8},
P. Calcidese\altaffilmark{9},
D. Carosati\altaffilmark{10}, 
P. Charlot\altaffilmark{11,12},
S. Ciprini\altaffilmark{13}, 
A. Di Paola\altaffilmark{14}, 
M. Dolci\altaffilmark{15},
N. V. Efimova\altaffilmark{4},
E. Forn\'e Scurrats\altaffilmark{16},
A. Frasca\altaffilmark{8}, 
A. C. Gupta\altaffilmark{17},
V. A. Hagen-Thorn\altaffilmark{18,19}, 
J. Heidt\altaffilmark{20},
D. Hiriart\altaffilmark{21},
T. S. Konstantinova\altaffilmark{18},
E. N. Kopatskaya\altaffilmark{18},
A. L\"ahteenm\"aki\altaffilmark{22}, 
L. Lanteri\altaffilmark{23}, 
V. M. Larionov\altaffilmark{18,19}, 
J.-F. Le Campion\altaffilmark{11,12},
P. Leto\altaffilmark{24},
E. Lindfors\altaffilmark{7},
E. Marilli\altaffilmark{8}
B. Mihov\altaffilmark{6},
E. Nieppola\altaffilmark{22},
K. Nilsson\altaffilmark{7},
J. M. Ohlert\altaffilmark{25}
E. Ovcharov\altaffilmark{26},
P. P\"a\"akk\"onen\altaffilmark{27},
M. Pasanen\altaffilmark{7},
B. Ragozzine\altaffilmark{1}, 
C. M. Raiteri\altaffilmark{23}, 
J. A. Ros\altaffilmark{28},
A. Sadun\altaffilmark{3},
A. Sanchez\altaffilmark{29},
E. Semkov\altaffilmark{6},
M. Sorcia\altaffilmark{32},
A. Strigachev\altaffilmark{6},
L. Takalo\altaffilmark{7},
M. Tornikoski\altaffilmark{22},
C. Trigilio\altaffilmark{8},
G. Umana\altaffilmark{8},
A. Valcheva\altaffilmark{6},
M. Villata\altaffilmark{23}, 
A. Volvach\altaffilmark{30},
J.-H. Wu\altaffilmark{31},
X. Zhou\altaffilmark{31}
}

\altaffiltext{1}{Astrophysical Institute, Department of Physics and Astronomy, \\
Clippinger 339, Ohio University, Athens, OH 45701, USA}
\altaffiltext{2}{Department of Astronomy, University of Michigan, 830 Dennison Building, \\
Ann Arbor, MI 48109-1042, USA}
\altaffiltext{3}{Department of Physics, University of Colorado Denver,
Campus Box 157, P. O. Box 173364, Denver, CO 80217-3364, USA}
\altaffiltext{4}{Pulkovo Observatory, Pulkovskoye Shosse, 65, 196140, St. Petersburg, Russia}
\altaffiltext{5}{Max-Planck-Institut f\"ur Radioastronomie, Radioobservatorium
Effelsberg, Max-Planck-Stra\ss e 28, D-53902 Bad M\"unstereifel-Effelsberg, Germany}
\altaffiltext{6}{Institute of Astronomy, Bulgarian Academy of Sciences,
72 Tsarigradsko shosse Blvd., 1784 Sofia, Bulgaria}
\altaffiltext{7}{Tuorla Observatory, University of Turku, 21500 Piikki\"o, Finland}
\altaffiltext{8}{Osservatorio Astrofisico di Catania, Viale A.\ Doria 6, \\
I-95125 Catania, Italy}
\altaffiltext{9}{Osservatorio Astronomico della Regione Autonoma Valle
d'Aosta, Italy}
\altaffiltext{10}{Osservatorio di Armenzano, Assisi, Italy}
\altaffiltext{11}{Universit\'e de Bordeaux, Observatoire Aquitain des Sciences de l'Univers, Floirac, France}
\altaffiltext{12}{CNRS, Laboratoire d'Astrophysique de Bordeaux -- UMR 5804, Floirac, France}
\altaffiltext{13}{Osservatorio Astronomico, Universit\`a di Perugia, Via B.\ Bonfigli, \\
I-06126 Perugia, Italy}
\altaffiltext{14}{Istituto Nazionale di Astrofisica (INAF), Osservatorio Astronomico di Roma,\\
Via Frascati, Monteporzio Cantone, Roma, Italy}
\altaffiltext{15}{Istituto Nazionale di Astrofisica (INAF), Osservatorio Astronomico di Teramo,\\
Via Maggini, Teramo, Italy}
\altaffiltext{16}{Code MPC B29, Astronomical Observatory L'Ampolla, Tarragona, Spain}
\altaffiltext{17}{Aryabhatta Research Institute of Observational Sciences
(ARIES), Manora Peak, Nainital -- 263 129, India}
\altaffiltext{18}{Astronomical Institute, St. Petersburg State University, \\
Universitetsky pr.\ 28, Petrodvoretz, 198504 St. Petersburg, Russia}
\altaffiltext{19}{Isaac Newton Institute of Chile, St. Petersburg Branch, \\
198504 St. Petersburg, Russia}
\altaffiltext{20}{Landessternwarte Heidelberg-K\"onigstuhl, K\"onigstuhl, \\
D-69117 Heidelberg, Germany}
\altaffiltext{21}{Instituto de Astronom\'{\i}a, Universidad Nacional
Aut\'onoma de M\'exico,\\
Apdo.\ Postal 877, 22800 Ensenada, B.C., M\'exico}
\altaffiltext{22}{Mets\"ahovi Radio Observatory, Helsinki University of Technology TKK, \\
Mets\"ahovintie 114, 02540 Kylm\"al\"a, Finland}
\altaffiltext{23}{Istituto Nazionale di Astrofisica (INAF), Osservatorio Astronomico di Torino,\\
Via Osservatorio 20, I-10025 Pino Torinese, Italy}
\altaffiltext{24}{INAF, Instituto di Radioastronomia, Sezione di Noto, Italy}
\altaffiltext{25}{Michael Adrian Observatory, Astronomie-Stiftung Trebur, Fichtenstra\ss e 7,
D-65468 Trebur, Germany}
\altaffiltext{26}{Sofia University ``St. Kliment Ohridski'', Faculty of Physics,
5 James Bourchier Blvd., 1164 Sofia, Bulgaria}
\altaffiltext{27}{Department of Physics and Mathematics, University of Joensuu, 
Joensuu, Finland}
\altaffiltext{28}{Agrupaci\`on Astronomica de Sabadell, Sabadell 08200, Spain}
\altaffiltext{29}{Gualba Observatory, MPC 442, Barcelona, Spain}
\altaffiltext{30}{Crimean Astrophysical Observatory, Naunchy, Crimea 98409, Ukraine}
\altaffiltext{31}{National Astronomical Observatories, Chinese Academy of Sciences,
20A Datun Road, Beijing 100012, China}
\altaffiltext{32}{Instituto de Astronom\'{\i}a, Universidad Nacional
Aut\'onoma de M\'exico,\\ 
Apdo. Postal 70-264, 04510 M\'exico, D.F., M\'exico}

\begin{abstract}
Prompted by a high optical state in September 2007, the Whole Earth
Blazar Telescope (WEBT) consortium organized an intensive optical,
near-IR (JHK) and radio observing campaign on the intermediate BL~Lac 
object 3C~66A throughout the fall and winter of 2007 -- 2008. In this 
paper, we present data from 28 observatories in 12 countries, covering 
the observing season from late July 2007 through February 2008. 
 
The source remained in a high optical state throughout the observing
period and exhibited several bright flares on time scales of $\sim 10$~days.
This included an exceptional outburst around September 15 -- 20, 2007, reaching
a peak brightness at $R \sim 13.4$. Our campaign revealed microvariability 
with flux changes up to $\vert {\rm d}R/{\rm d}t \vert \sim 0.02$~mag/hr. 

Our observations do not reveal evidence for systematic spectral variability
in the overall high state covered by our campaign, in agreement with
previous results. In particular, we do not find evidence for spectral 
hysteresis in 3C~66A for which hints were found in an earlier campaign
in a somewhat lower flux state. We did also not find any evidence for
spectral lags in the discrete correlation functions between different
optical bands. 

We infer a value of the magnetic field in the emission region of $B \sim
19 \, e_B^{2/7} \, \tau_h^{-6/7} \, D_1^{13/7}$~G, where $e_B$ is the
magnetic field equipartition fraction, $\tau_h$ is the shortest observed
variability time scale in units of hours, and $D_1$ is the Doppler factor
in units of 10. From the lack of systematic spectral variability, we can
derive an upper limit on the Doppler factor, $D \le 28 \, \tau_h^{-1/8} 
\, e_B^{3/16}$. This is in perfect agreement with superluminal motion
measurements with the VLBI/VLBA of $\beta_{\rm app} \le 27$ and 
argues against models with very high Lorentz factors of $\Gamma \gtrsim 
50$, required for a one-zone synchrotron-self-Compton interpretation of
some high-frequency-peaked BL Lac objects detected at TeV $\gamma$-ray 
energies. 
\end{abstract}

\keywords{galaxies: active --- BL Lacertae objects: individual (3C~66A) 
--- gamma-rays: theory --- radiation mechanisms: non-thermal}  

\section{Introduction}

Blazars are the most violent class of active galactic nuclei, consisting
of flat-spectrum radio quasars (FSRQs) and BL~Lac objects. They exhibit
rapid variability down to time scales as short as a few minutes 
\citep{aharonian07,albert07b}. Their observed flux is dominated by a
non-thermal continuum exhibiting two broad spectral bumps: A low-frequency
bump from radio to UV -- X-ray frequencies, and a high-frequency component
from X-ray to $\gamma$-rays. 

In the framework of relativistic jet models, the low-frequency (radio
-- optical/UV) emission from blazars is interpreted as synchrotron
emission from nonthermal electrons in a relativistic jet. The
high-frequency (X-ray -- $\gamma$-ray) emission could either be
produced via Compton upscattering of low frequency radiation by the
same electrons responsible for the synchrotron emission \citep[leptonic
jet models; for a recent review see, e.g.,][]{boettcher07}, or 
due to hadronic processes initiated by relativistic protons 
co-accelerated with the electrons \citep[hadronic models, for 
a recent discussion see, e.g.,][]{muecke01,muecke03}. 

The blazar 3C~66A (= 0219+428) is classified as a low-frequency 
peaked BL~Lac object (LBL), a class also commonly referred to 
as radio selected BL~Lac objects. Its nonthermal low-frequency 
spectral component extends from radio frequencies through soft X-rays 
and typically peaks in the optical frequency range. The high-frequency 
component seems to peak in the multi-MeV -- GeV energy range. Since its 
optical identification by \cite{wills74}, 3C~66A has been the target of 
many radio, IR, optical, X-ray, and $\gamma$-ray observations in the past, 
although it is not as regularly monitored at radio frequencies as many other 
blazars due to problems with source confusion with the nearby radio 
galaxy 3C~66B (6'.5 from 3C~66A), in particular at lower (4.8 and 
8~GHz) frequencies \citep{aller94,takalo96}.

To date, about 2 dozen blazars have been detected at very high energies 
($> 100$~GeV) with ground-based air \v Cerenkov telescope facilities. Until very 
recently, all TeV blazars belonged to the sub-class of high-frequency peaked 
BL~Lac objects (HBLs). However, the recent detections of the intermediate
BL~Lac object W~Comae \citep{beilicke08}, the low-frequency peaked BL~Lac
object BL~Lacertae \citep{albert07a}, and even the flat-spectrum radio
quasar 3C~279 \citep{albert08} demonstrate the potential to extend the 
extragalactic TeV source list to all classes of blazars, in particular
intermediate and low-frequency peaked BL~Lac objects (LBLs) with lower 
$\nu F_{\nu}$ peak frequencies in their broadband spectral energy 
distributions (SEDs). 

3C~66A has been suggested for quite some time as a promising candidate 
for detection by the new generation of atmospheric \v Cerenkov 
telescope facilities like MAGIC or VERITAS \citep[e.g.][]{cg02}
(it is too far north in the sky to be observed by HESS), 
and intensive
observations of 3C~66A by the VERITAS collaboration have recently
resulted in the significant detection of VHE $\gamma$-ray emission
from 3C~66A in September -- October 2008 \citep{swordy08}.

In the optical,
3C~66A is generally observed as a point source, with no indication of
the host galaxy. The host galaxy of 3C~66A was marginally resolved by
\cite{wurtz96}. They found $R_{\rm Gunn} = 19.0^{\rm mag}$ for the host 
galaxy; the Hubble type could not be determined. In 3C~66A, a weak Mg~II 
emission line has been detected by \cite{miller78}. This led 
to the determination of its redshift at $z = 0.444$, which was 
later confirmed by the detection of a weak Ly$\alpha$ line in 
the IUE spectrum of 3C~66A \citep{lanzetta93}. However, as recently 
pointed out by \cite{bramel05}, these redshift determinations are
actually still quite uncertain \citep[see also][]{finke08a}. In this 
paper, we do base our analysis on a redshift value of $z = 0.444$, but 
remind the reader that some results of the physical interpretation should 
be considered as tentative pending a more solid redshift determination. 

The long-term variability of 3C~66A at near-infrared (J, H, and K bands) 
and optical (U, B, V, R, I) wavelengths has recently been compiled and
analyzed by \cite{fl99} and \cite{fl00}, respectively. Variability at
those wavelengths is typically characterized by variations over $\ls
1.5$~mag on time scales ranging from $\sim 1$~week to several years.
A positive correlation between the B -- R color (spectral hardness) 
and the R magnitude has been found by \cite{vagnetti03}. The most recent
multiwavelength campaign on 3C~66A in 2003 -- 2004 \citep{boettcher05} 
found hints for spectral hysteresis, with the B -- R hardness peaking
several days prior to the R- and B-band fluxes during large flares. 
That campaign also confirmed the results from an intensive long-monitoring 
effort by \cite{takalo96}, revealing evidence for rapid microvariability, 
including a decline $\sim 0.2$~mag within $\sim 6$~hr. \cite{lainela99} 
also report a 65-day periodicity of the source in its optically bright 
state, which has so far not been confirmed in any other analysis. However,
the campaign of \cite{boettcher05} showed several major outbursts separated
by $\sim 50$ -- 57~days, possibly also indicating a quasi-periodic behavior.

Superluminal motion of individual radio components of the jet
has been detected by \cite{jorstad01}. While the identification
of radio knots across different observing epochs is not unique,
\cite{jorstad01} favor an interpretation implying superluminal
motions of up to $\beta_{\rm app} \sim 19 \, h^{-1} \approx 
27$, where $h = H_0 / (100$~km~s$^{-1}$~Mpc$^{-1}$) parameterizes
the Hubble constant.
This would imply a lower limit on the bulk Lorentz factor
of the radio emitting regions of $\Gamma \ge 27$. Theoretical
modeling using a time-dependent leptonic jet model produced an 
acceptable fit to the SED and light curve of 3C~66A during the 
2003 -- 2004 campaign with such a choice of $\Gamma$ \citep{joshi07}.

In the fall of 2007, 3C~66A was found in a very active state, reaching
a peak brightness around Sept. 14 of $R \sim 13.4$. This triggered a
new optical -- IR -- radio observing campaign by the Whole Earth Blazar 
Telescope (WEBT\footnote{\tt http://www.oato.inaf.it/blazars/webt/}) 
collaboration with intensive monitoring throughout fall and winter 
2007 -- 2008. In this paper we present collected data from late 
July 2007 through early March 2008. This high optical state also 
triggered very-high-energy $\gamma$-ray observations with the 
VERITAS array. Continued observations by VERITAS throughout 
the observing season in 2008 resulted in the VHE $\gamma$-ray detection 
mentioned above \citep{swordy08}.
Details of those observations will be published
in a separate paper. In the following, we will present in \S \ref{observations}
a summary of the observations and data analysis, and present light curves
in the various frequency bands. In \S \ref{variability} we test for spectral
variability to be derived from our results. Cross-correlations between
the variability in different optical bands are presented in \S 
\ref{crosscorrelations}. We discuss our results and derive limits on 
the magnetic field and the Doppler factor in \S \ref{discussion},
and we summarize in \S \ref{summary}.

Throughout this paper, we refer to $\alpha$ as the energy 
spectral index, $F_{\nu}$~[Jy]~$\propto \nu^{-\alpha}$. A 
cosmology with $\Omega_m = 0.3$, $\Omega_{\Lambda} = 0.7$, 
and $H_0 = 70$~km~s$^{-1}$~Mpc$^{-1}$ is used. In this cosmology,
and using the redshift of $z = 0.444$, the luminosity distance 
of 3C~66A is $d_L = 2.46$~Gpc. 

\begin{figure}
\plotone{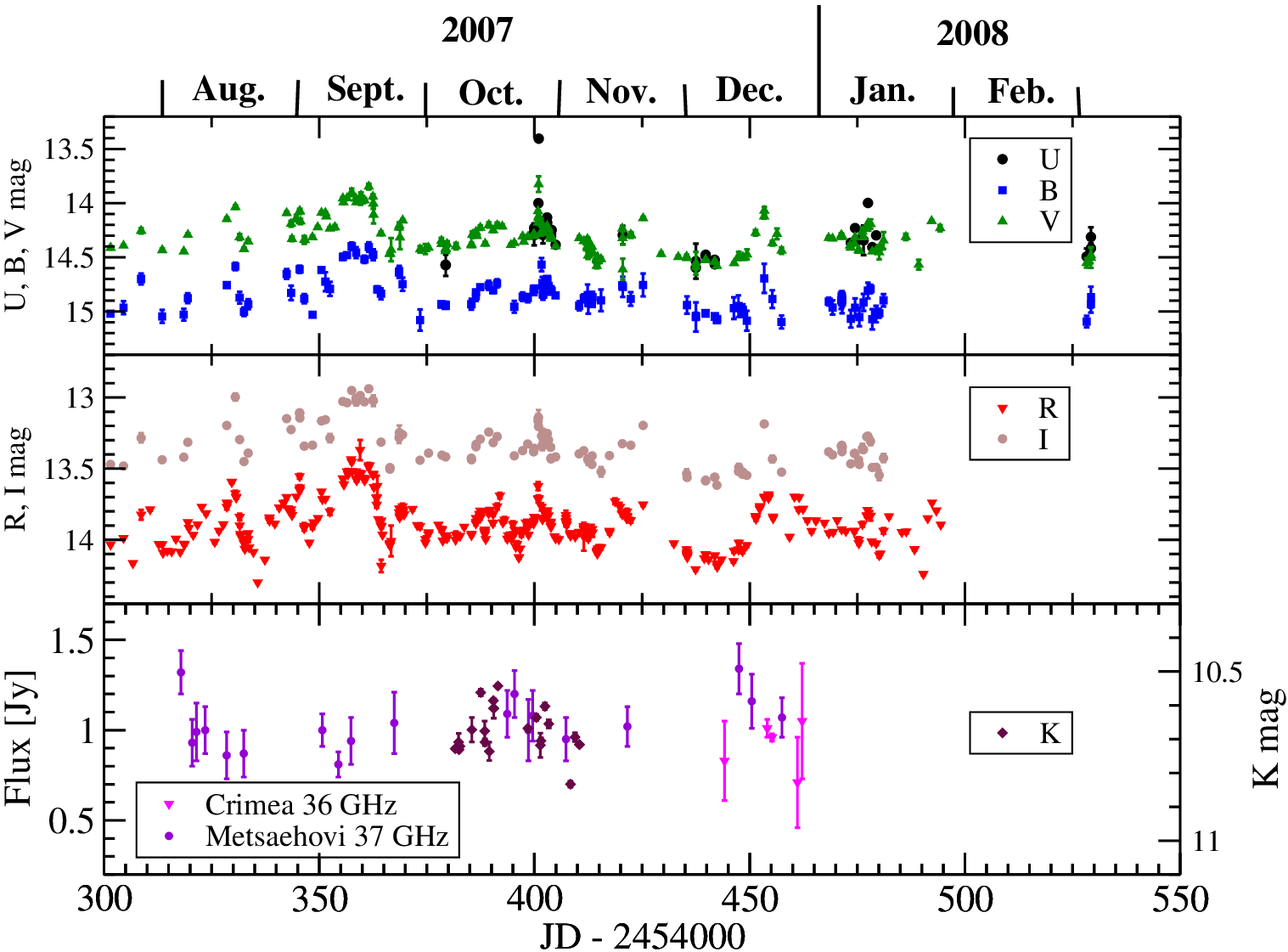}
\caption{Timeline of the WEBT campaign on 3C~66A in 2007 -- 2008, 
including the observed (before extinction correction) optical (UBVRI), 
near-infrared K-band, and radio light curves}
\label{timeline}
\end{figure}

\section{\label{observations}Observations, data reduction, 
and light curves}

3C~66A was observed in a coordinated multiwavelength campaign at radio, 
near-IR, and optical frequencies by the WEBT collaboration during the
2007 -- 2008 observing season, from late July 2007 through early
March 2008. The object is being continuously monitored at radio 
to optical wavelengths by the GLAST-AGILE Support Program \citep[GASP;
see][]{villata08}, whose observers contributed also to this campaign. 
The overall timeline of the campaign, along with the measured long-term 
light curves at radio, infrared, and optical frequencies is 
illustrated in Fig. \ref{timeline}. Table \ref{observatories} 
lists all participating observatories which contributed data to 
this campaign. In this section, we will describe the individual 
observations in the various frequency ranges and outline the data 
reduction and analysis.

\begin{deluxetable}{lccc}
\tabletypesize{\scriptsize}
\tablecaption{List of observatories that contributed data to this campaign. 
In addition to the standard optical Johnson's-Cousin's UBVRIJHK bands, we 
obtained optical/near-IR data in the c, i, and o bands. The central wavelengths 
of these bands are c = 4206 \AA, i = 6658 \AA, and o = 9173 \AA.}
\tablewidth{0pt}
\tablehead{
\colhead{Observatory} & \colhead{Specifications} & \colhead{frequency / filters / energy range} &
\colhead{$N_{\rm obs}$}
}
\startdata
\multispan 4 \hss \bf Radio Observatories \hss \\
\noalign{\smallskip\hrule\smallskip}
Crimean Radio Obs., Ukraine   & 22 m  & 36 GHz           & 5 \\
Medicina, Italy               & 32 m  & 5, 8, 22 GHz     & 19 \\
Mets\"ahovi, Finland          & 14 m  & 37 GHz           & 32 \\
Noto, Italy                   & 32 m  & 43 GHz           & 19 \\
UMRAO, Michigan, USA          & 26 m  & 4.8, 8, 14.5 GHz & 12  \\
\noalign{\smallskip\hrule\smallskip}
\multispan 4 \hss \bf Infrared Observatories \hss \\
\noalign{\smallskip\hrule\smallskip}
Campo Imperatore, Italy     & 1.1 m & J, H, K  & 72 \\
\noalign{\smallskip\hrule\smallskip}
\multispan 4 \hss \bf Optical Observatories \hss \\
\noalign{\smallskip\hrule\smallskip}
ARIES, Nainital, India        & 104 cm        & B, V, R, I    & 15 \\
Armenzano, Italy              & 35, 40 cm     & B, V, R, I    & 70 \\
Belogradchik, Bulgaria        & 60 cm         & V, R, I       & 17 \\
Obs. de Bordeaux, France      & 20 cm         & V             & 34 \\
Catania, Italy                & 91 cm         & U, B, V       & 60 \\
Crimean Astr. Obs., Ukraine   & 70 cm         & B, V, R, I    & 147 \\
Gualba Obs., Spain            & 35 cm         & V, R, I       & 12 \\
Jakokoski Obs., Finland       & 51 cm         & R             & 69 \\
Kitt Peak (MDM), Arizona, USA & 130 cm        & U, B, V, R, I & 333 \\
L'Ampolla, Spain              & 36 cm         & R             & 396 \\
Michael Adrian Obs., Germany  & 120 cm        & R             & 180 \\
New Mexico Skies Obs., USA    & 30 cm         & V, R, I       & 18 \\
Perugia, Italy                & 40 cm         & R             & 25 \\
Roque (KVA), Canary Islands   & 35 cm         & R             & 318 \\
Rozhen, Bulgaria              & 200 cm        & U, B, V, R, I & 139 \\
Sabadell, Spain               & 50 cm         & V, R          & 464 \\
San Pedro M\'artir, Mexico    & 84 cm        & R             & 10 \\
St. Petersburg, Russia        & 38 cm         & B, V,. R, I   & 50 \\
Torino, Italy                 & 105 cm        & B, V, R, I    & 77 \\
Tuorla, Finland               & 103 cm        & R             & 19 \\
Valle d'Aosta, Italy          & 81 cm         & B, V, R, I    & 53 \\
Xinglong, China               & 60 cm         & c, i, o       & 699 \\
\noalign{\smallskip\hrule\smallskip}
\enddata
\label{observatories}
\end{deluxetable}

\subsection{\label{optical}Optical and infrared observations}

In the optical component of the WEBT campaign, 22
observatories in 12 countries contributed 3461 individual 
photometric data points. The observing strategy and data analysis 
followed to a large extent the standard procedure used for the 
previous, successful WEBT campaign on 3C~66A in 2003 -- 2004
\citep{boettcher05}. 
Observers were asked to perform bias (and, where necessary, dark) 
corrections as well as flat-fielding on their frames, and obtain 
instrumental magnitudes, applying either aperture photometry (using 
IRAF or CCDPHOT) or Gaussian fitting for the source 3C~66A and the 
comparison stars no. 13, 14, 21, and 23 in the tables of \cite{gp01}, 
where high-precision standard magnitudes for these stars have been 
published. This calibration has then been used to convert instrumental 
to standard photometric magnitudes for each data set. In the next step, 
unreliable data points (with large error bars at times when higher-quality 
data points were available) were discarded. Our data did not provide 
evidence for significant variability on sub-hour time scales. Consequently, 
error bars on individual data sets could be further reduced by re-binning 
on time scales of 20~min. 
Finally, there may be systematic offsets between different instruments 
and telescopes. Wherever our data sets contained sufficient independent
measurements to clearly identify such offsets, individual data sets 
were corrected by applying appropriate correction factors. The resulting 
offsets, by which individual data sets have been corrected to create a
uniform light curve, are listed in Tab. \ref{offsets}.

\begin{figure}
\plotone{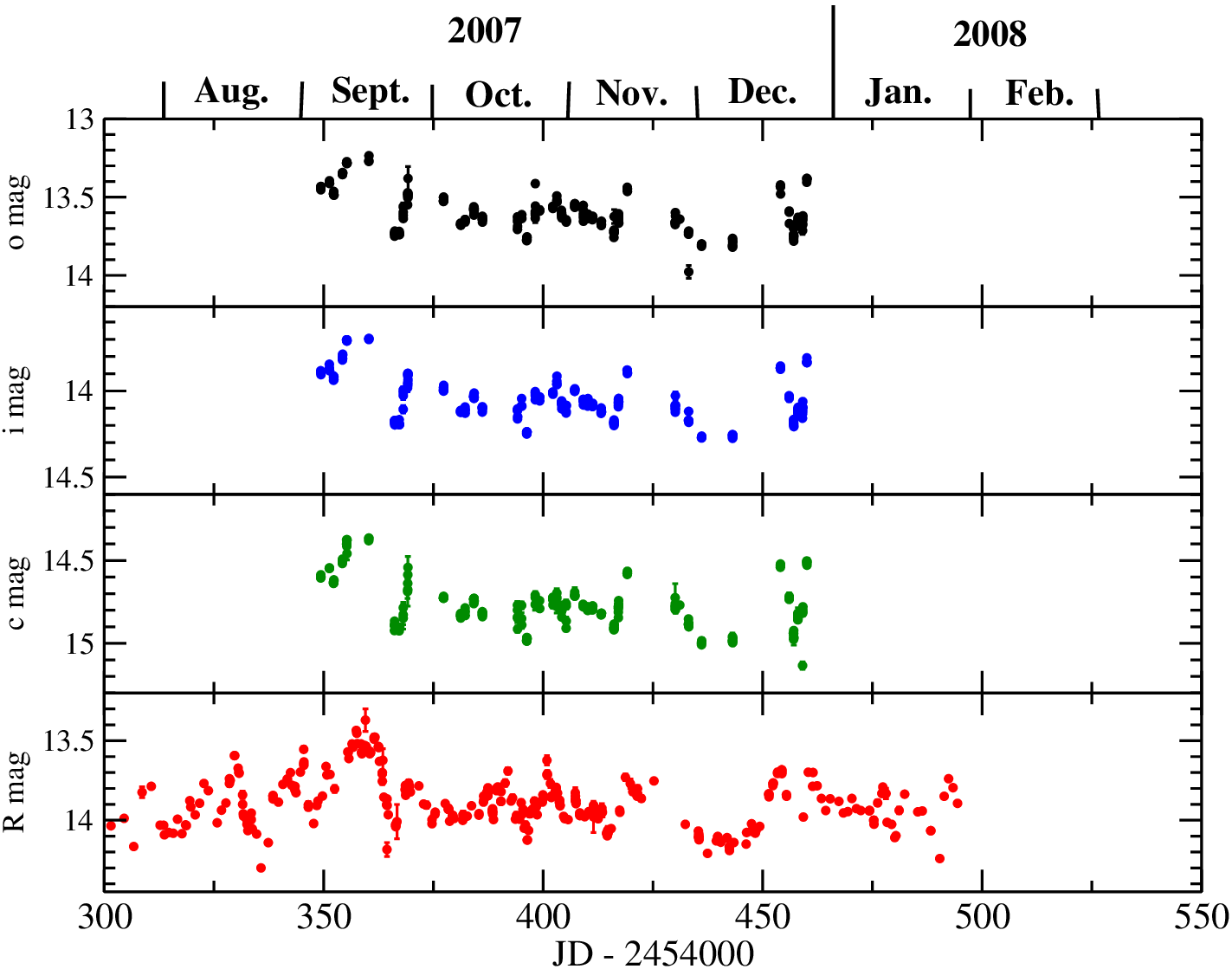}
\caption{Observed light curves in the optical R, c, i, and o bands (before
extinction correction) over the entire duration of the campaign. The central
wavelengths of these bands are c = 4206 \AA, i = 6658 \AA, and o = 9173 \AA.}
\label{rcio}
\end{figure}

In order to provide information on the intrinsic broadband 
spectral shape (and, in particular, a reliable extraction 
of B - R color indices), the data were de-reddened. For 
this purpose, Galactic Extinction coefficients were 
calculated using Table 3 of \cite{cardelli89}, based on $A_B 
= 0.363$~mag\footnote{\tt http://nedwww.ipac.caltech.edu/}
and $R_V = 3.1$. As mentioned in the introduction, the R 
magnitude of the host galaxy of 3C~66A is $\sim 19$~mag, so 
its contribution is negligible compared to the average AGN
magnitude of $R \ls 14$, and no host-galaxy correction 
was applied.

\begin{deluxetable}{ccccc}
\tabletypesize{\scriptsize}
\tablecaption{Inter-instrumental correction offsets for optical data. Only
those telescopes are listed for which non-zero offsets were required.}
\tablewidth{0pt}
\tablehead{
\colhead{Observatory} & \colhead{B} & \colhead{V} & \colhead{R} & \colhead{I} 
}
\startdata
ARIES&			+0.12&	-&	-&	-\\
Belogradchik&		-&	-&	-&	+0.03\\
Rozhen&                 +0.06&  -&	-&	-\\
Torino&			-0.12&	-&	-&	-\\
Valle d'Aosta&          -&      -&	-&	+0.035\\
\enddata
\label{offsets}
\end{deluxetable}

\begin{figure}
\plotone{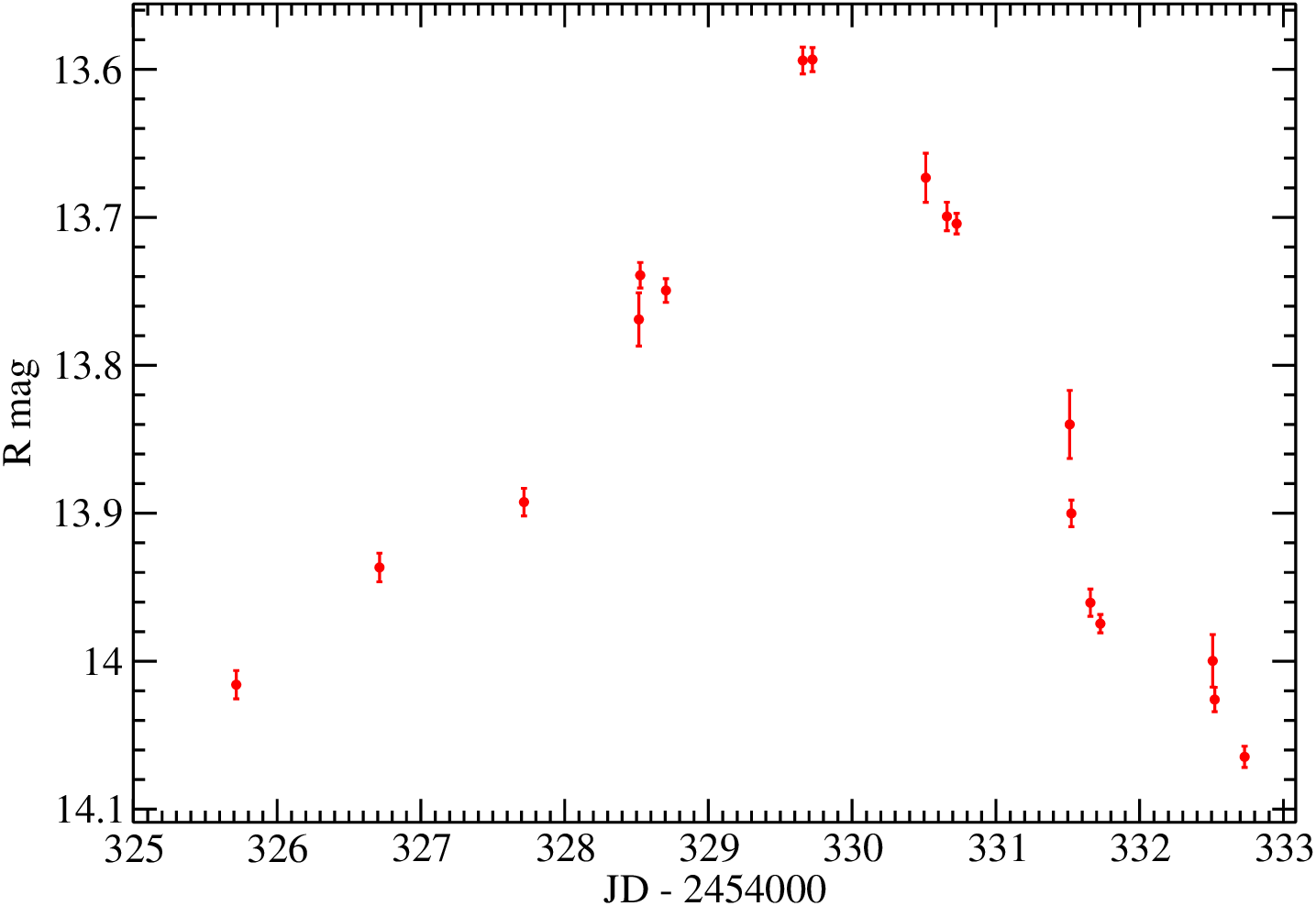}
\caption{Details of the R-band light curve (no extinction correction) 
during the outburst around August 16, 2007. Particularly remarkable 
is the rapid decline on JD 2454331 (August 18) with a slope of 
$\sim 0.016$~mag/hr. }
\label{R330}
\end{figure}

As a consequence of the chosen observing strategy, the 
R- and B-band light curves are the most densely sampled 
ones. The R-band light curve over the entire duration of the 
campaign is compared to the light curves at all other optical 
bands as well as the near-IR K-band and 36/37~GHz radio light
curves in Figs.~\ref{timeline} and \ref{rcio}. 
These figures illustrate that the object remained in a very bright
and active state, with an average magnitude of $R \sim 14.0$, 
throughout the campaign. The source exhibited several large flares
with $\Delta R \gs 0.5$~mag out to $R \sim 13.5$ -- $13.6$, on 
characteristic time scales of $\sim 10$~days. The most dramatic 
outburst occurred around September 14, with a peak magnitude of 
$R \sim 13.4$. As already found in 2003 -- 2004 \citep{boettcher05}, 
visual inspection of individual major outbursts suggests periods of 
more rapid decline than rise (see, e.g., Fig. \ref{R330}).

\begin{figure}
\plotone{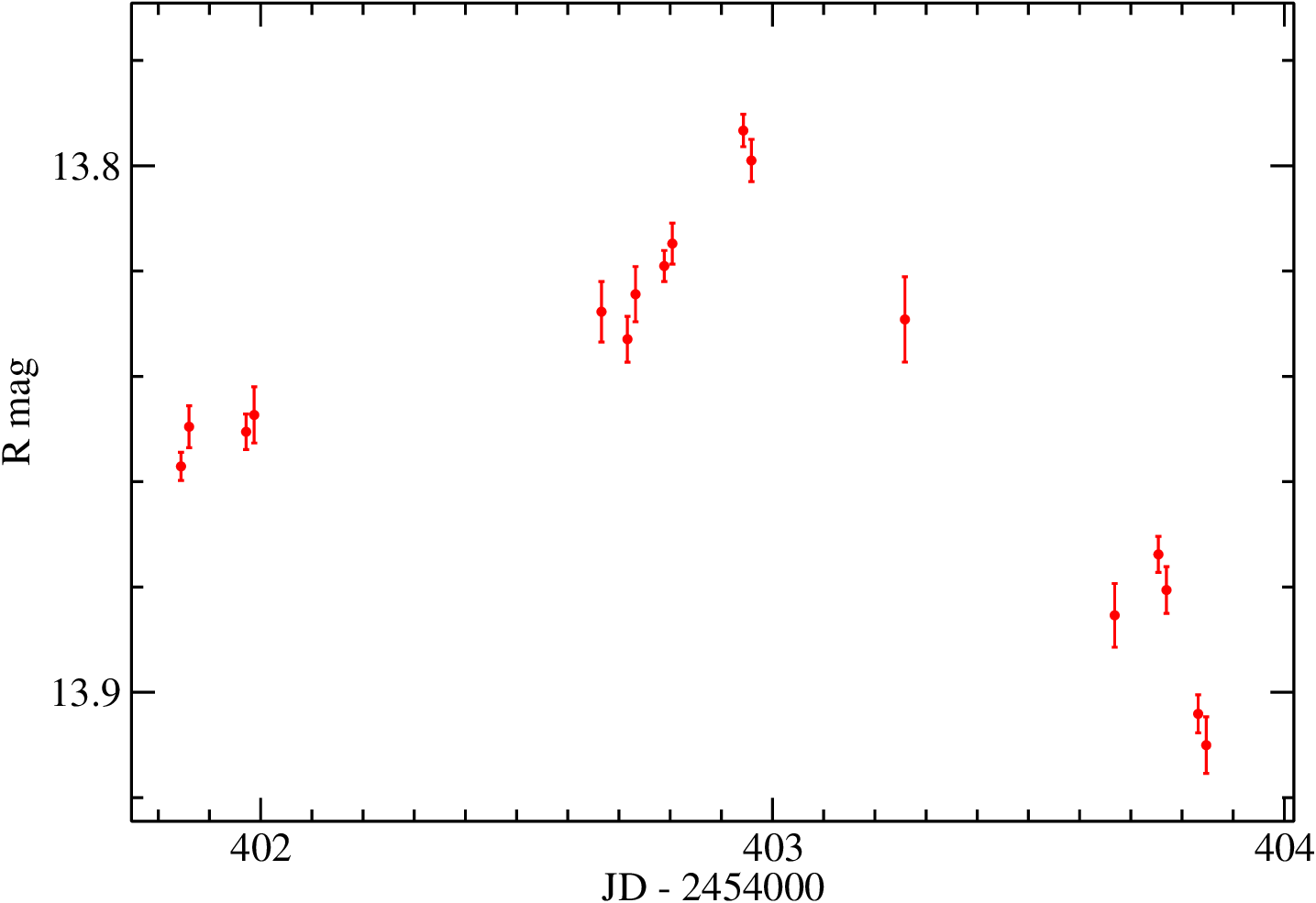}
\caption{Details of the R-band light curve (no extinction correction) on Oct. 
28 -- 29, 2007 (JD~2454402 -- 2454403). The rise around JD~2454402.8 has a 
slope of $\sim -0.008$~mag/hr.}
\label{R403}
\end{figure}

Our campaign revealed clear evidence for rapid intraday microvariability. Two 
examples are shown in Figures \ref{R330} and \ref{R403}, revealing a rapid decline
of the optical brightness by ${\rm d}R/{\rm d}t \sim 0.016$~mag/hr on JD~2454331 (August 18,
2007), and a rapid rise by ${\rm d}R/{\rm d}t \sim -0.008$~mag/hr on JD~2454402 (October 28, 
2007). Our results do not provide any evidence for periodicity or quasi-periodicity.
Overall, the variability patterns in all optical bands are very well correlated
with no discernable time lag between bands (see \S \ref{crosscorrelations}).

In the context of our WEBT campaign, 3C~66A was also observed at near-infrared 
wavelengths in the J, H, and K bands with the AZT-24 1.1-m telescope 
at Campo Imperatore, Italy. 
The primary data were analyzed using the same standard technique as the optical 
data (see above), including flat-field subtraction, extraction of instrumental 
magnitudes, calibration against comparison stars to obtain standard magnitudes, 
and de-reddening. The sampling was not dense enough to allow an improvement of 
the data quality by re-binning. Also, the IR observations covered only a period
of $\sim 1$~month so that we can not draw any conclusions concerning
correlated variability with optical or radio fluxes. 

\begin{figure}
\plotone{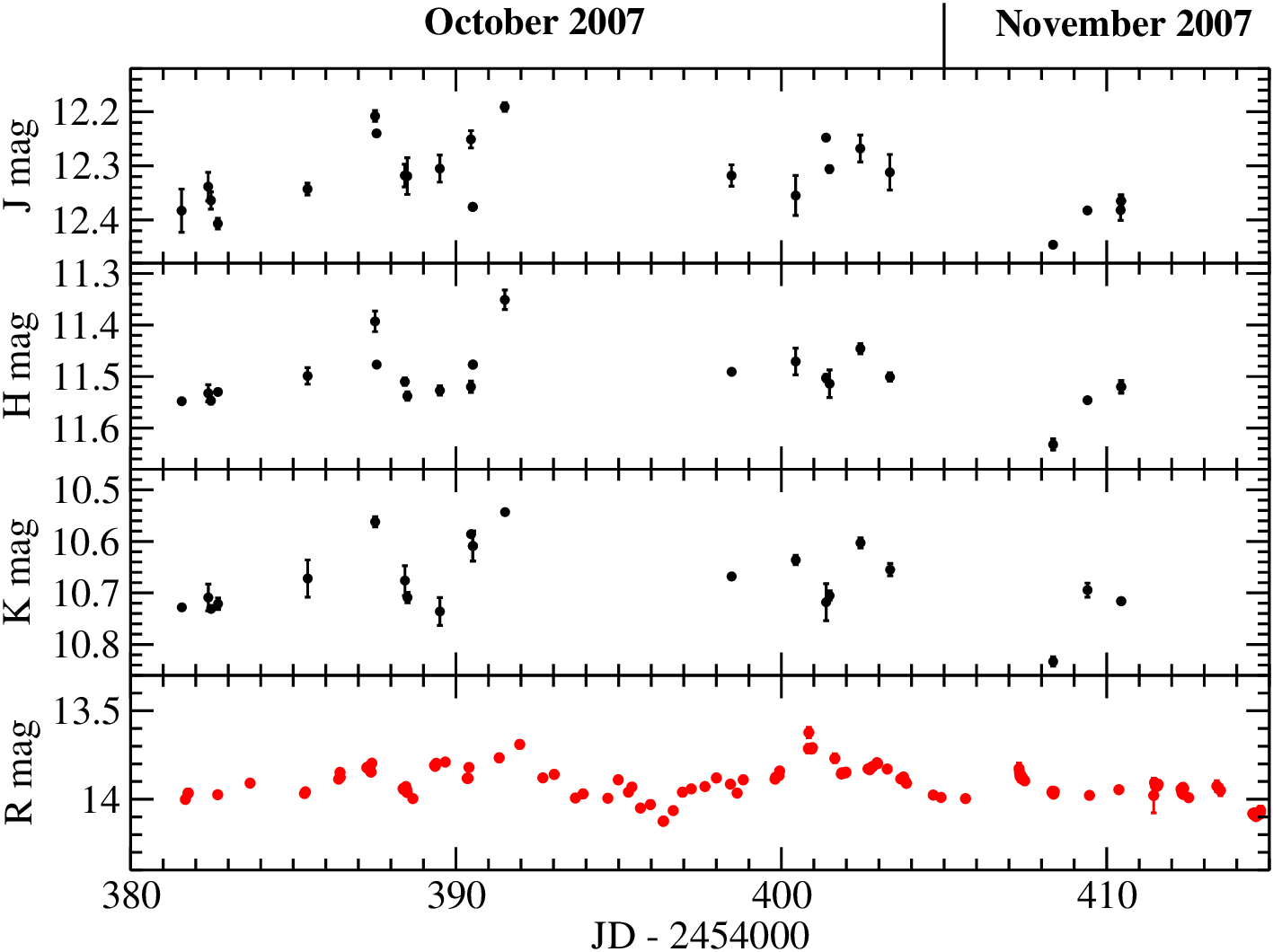}
\caption{Light curves in the near-infrared J, H, and K bands during the campaign,
compared to the R-band light curve.}
\label{IR_lc}
\end{figure}

The resulting K-band light curve is included in Fig. \ref{timeline}, and
a comparison of all three IR band light curves (J, H, K) is shown in 
Fig. \ref{IR_lc}. Within the limited time resolution and coverage, the
JHK variability tracks the variability in the optical bands. For example, 
small BVRI flares on JD~2454387 and JD~2454391 -- 2454392 are closely
matched by the JHK light curves. The JHK light curves exhibit variability
on similar time scales as the optical bands, but with somewhat smaller
amplitudes of $\Delta(JHK) \ls 0.3$~mag. Rapid variability with rise
and decay rates of $\Delta(JHK) \gs 0.1$~mag/day can be seen, in
particular around JD~2454387 -- 2454392.

\begin{figure}
\plotone{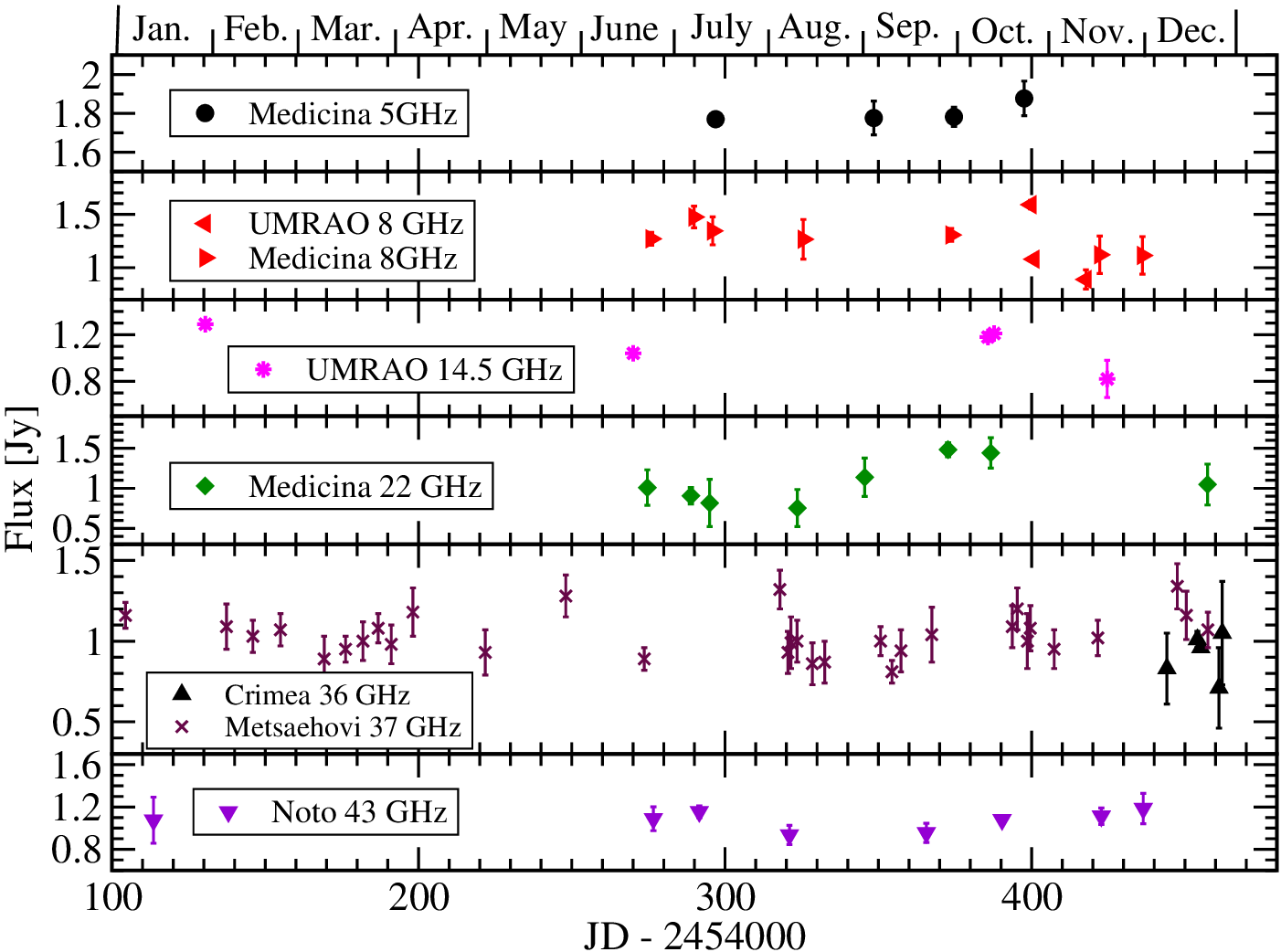}
\caption{Radio light curves of 3C~66A in 2007.}
\label{radio_lc}
\end{figure}

\subsection{\label{radio}Radio observations}

At radio frequencies, the object was monitored using the Noto radio 
telescope at 43~GHz, the 22~m radio telescope (RT~22) of the Crimean 
Radio Observatory at 36~GHz, the University of Michigan Radio Astronomy 
Observatory (UMRAO) 26~m telescope, at 4.8, 8, and 14.5~GHz, the 
14~m Mets\"ahovi Radio Telescope of the Helsinki University of 
Technology, at 37~GHz, and the Medicina radio telescope at 5, 8, 
and 22~GHz. 

The observations at 43 GHz have been performed using the 32-m radio
telescope in Noto (Italy). The observations reported in this paper 
were carried out using the 43 GHz supereterodyne cooled receiver
with a 400~MHz instantaneous band. For details of the data analysis 
and flux calibration, see \cite{ott94}.

The observations at 36~GHz were carried out at the 22~m Crimean 
Astrophysical Observatory radio telescope (RT~22), using two similar 
Dicke switched radiometers. For details of the analysis and 
calibration of data from the RT~22, see \cite{volvach06}.
At the UMRAO, the source was monitored in the course of the
on-going long-term blazar monitoring program. The data were
analyzed following the standard procedure described in
\cite{aller85}. As mentioned above, the sampling was rather 
poor, and some individual errors were rather large due to 
source confusion problems with 3C~66B. 

The 37 GHz observations were made with the 13.7 m diameter Mets\"ahovi 
radio telescope, which is a radome enclosed paraboloid antenna situated in
Finland (24$^o$ 23' 38''E, +60$^o$ 13' 05''). The measurements were made with 
a 1 GHz-band dual beam receiver centered at 36.8 GHz. A detailed description 
on the data reduction and analysis is given in \cite{ter98}. 

The radio data at 5, 8, and 22~GHz from the Medicina antenna were 
taken within a long-term monitoring program of gamma-ray bright 
blazars. For details on the analysis of data from the Medicina 
antenna see \cite{bach07}.

The 36 and 37~GHz light curves from the Crimean Astrophysical
Observatory and Mets\"ahovi, respectively, are included in
Fig. \ref{timeline} and in the compilation of all radio light
curves throughout 2007 in Fig. \ref{radio_lc}. They reveal 
small-amplitude ($\Delta F / F \ls 20$~\%) variability 
on time scales of $\sim 10$ -- 20~days. The sampling is 
insufficient to allow any conclusions about possible correlations
of the radio variability with optical or infrared variability.

The most dramatic radio variability appears to occur at the
highest observed radio frequency of 43~GHz at the Noto radio
observatory (see Fig. \ref{radio_lc}), possibly indicating
erratic variability on a one-day time scale. However, we point 
out that the observed erratic variability at higher radio frequencies 
($\ge 14.5$~GHz) may, at least in part, be a consequence of 
interstellar scintillation \citep[for a more detailed discussion of 
this aspect, see][]{boettcher05}.

\begin{figure}
\plotone{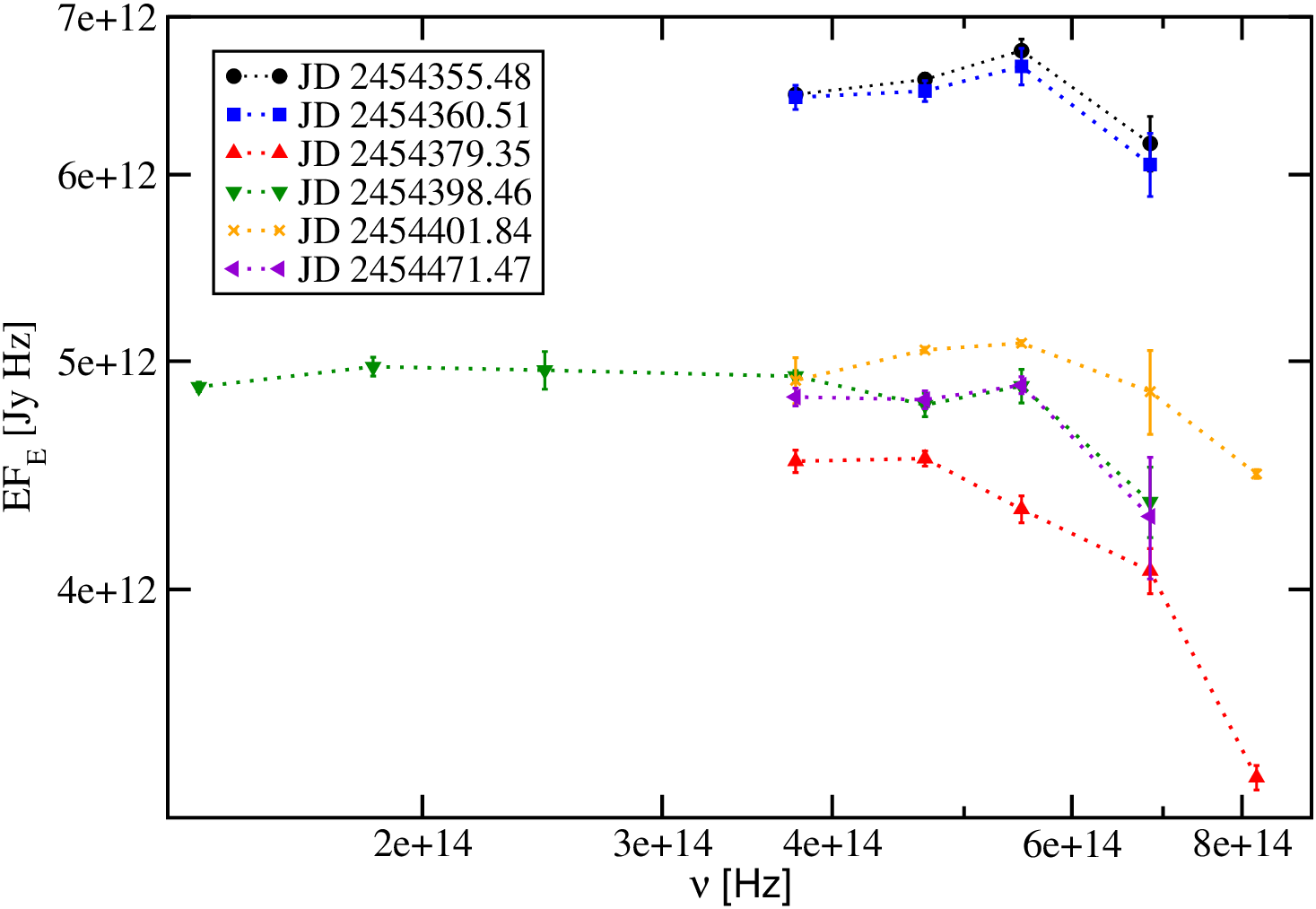}
\caption{Intrinsic (after extinction correction) optical -- near-IR 
spectral energy distributions during 6 epochs throughout the campaign. }
\label{SEDs}
\end{figure}

\begin{figure}
\plotone{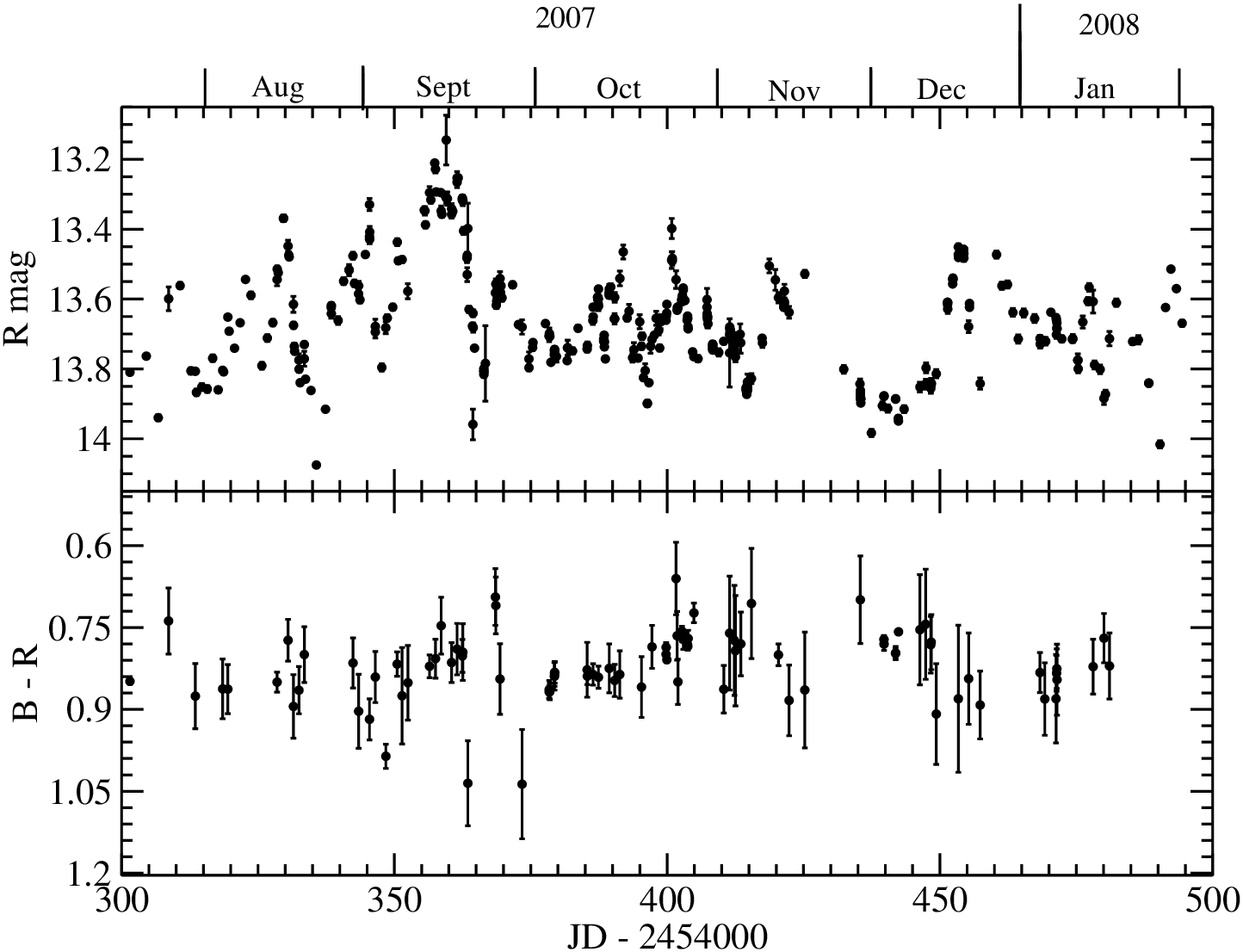}
\caption{Intrinsic (after extinction correction) light curves of the R 
magnitudes and the intrinsic B -- R color index of 3C~66A over the duration 
of the entire campaign. No obvious correlation can be identified. }
\label{B_R_lc}
\end{figure}

\section{\label{variability}Optical spectral variability}

To investigate optical spectral variability observed during our campaign,
we first computed snap-shot optical continuum spectra for several points
in time throughout the campaign. A compilation of several such optical
/ near-IR sections of the SEDs are plotted in Fig. \ref{SEDs}. As in
our earlier campaign on 3C~66A \citep{boettcher05}, we find the (presumably)
synchrotron peak of the SED in the optical regime, 
around (4 -- 6)$\times 10^{14}$~Hz. In order to investigate whether 
there is a systematic shift of the synchrotron peak with varying optical 
flux level, we calculated 
B - R color indices for any pair of B and R magnitude measurements taken 
within 20 minutes of each other. Fig. \ref{B_R_lc} shows the resulting 
plot of the B - R color index vs. time, together with the R band light curve. 
This figure does not confirm the hint in our compilation of snap-shot optical/IR 
SEDs of a correlation between the brightness and the hardness of the source 
during our campaign. 

\begin{figure}
\plotone{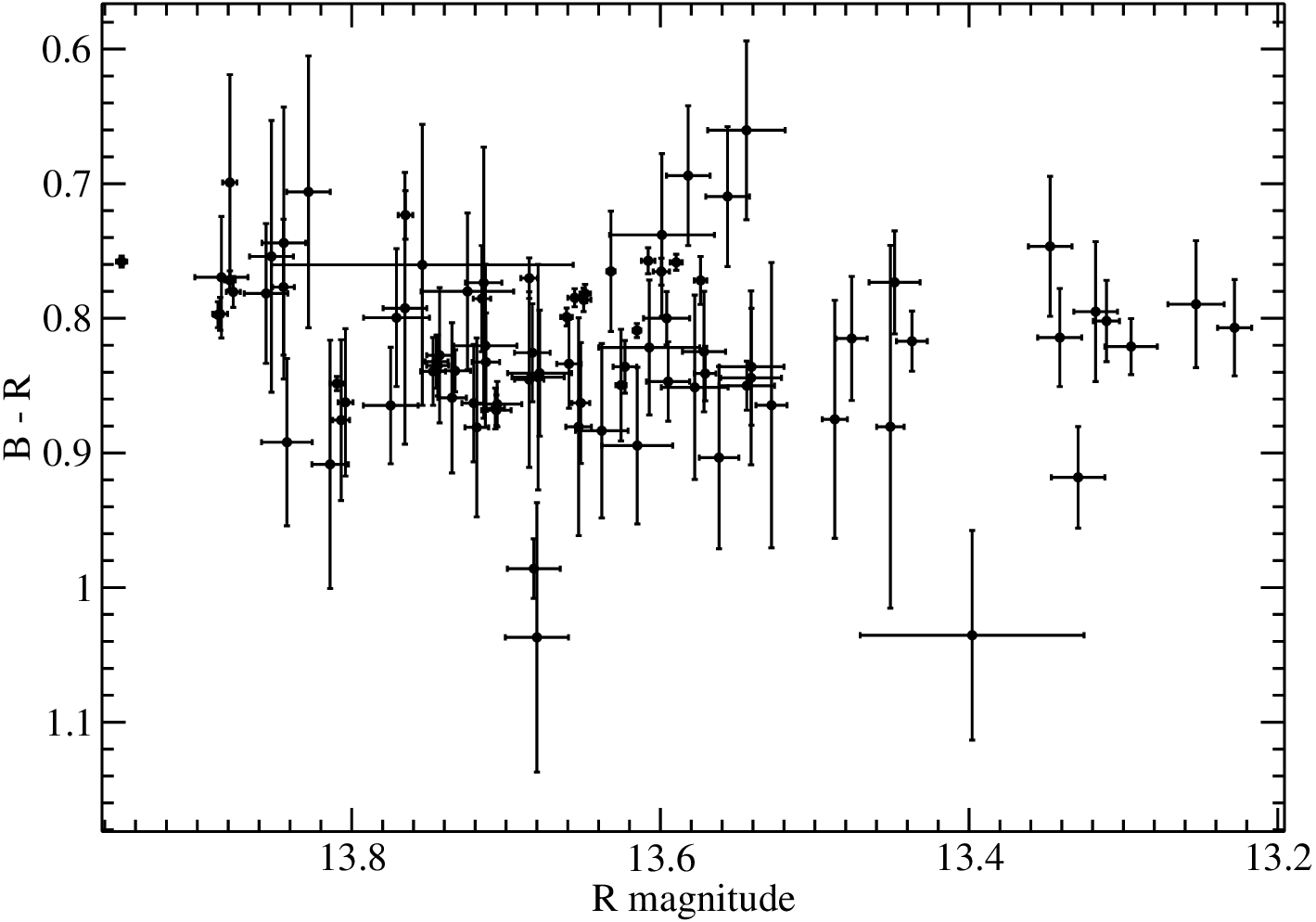}
\caption{Intrinsic optical hardness -- intensity diagram for the complete data 
set over the entire duration of the campaign. No correlation of the B - R
index with R-band magnitude can be identified. }
\label{HID}
\end{figure}

This is further illustrated in Fig. \ref{HID}, showing the hardness-intensity
diagram of B - R color index vs. R-band magnitude. No trend of B - R index
with brightness can be found. In our previous campaign \citep{boettcher05},
we had found a weak positive correlation of hardness with flux for low
flux states, $R_{\rm dered} \gs 14.0$ \citep[in agreement with the 
positive correlation identified by][]{vagnetti03}, but no correlation at 
brighter flux states. Similar results were also found for the low-frequency
peaked BL~Lac object BL~Lacertae \citep{papadakis07}. Since during the entire 
2007-2008 campaign the source remained in a bright state with $R_{\rm dered} 
\ls 14.0$, our results are in perfect agreement with our previous findings. 

\begin{figure}
\plotone{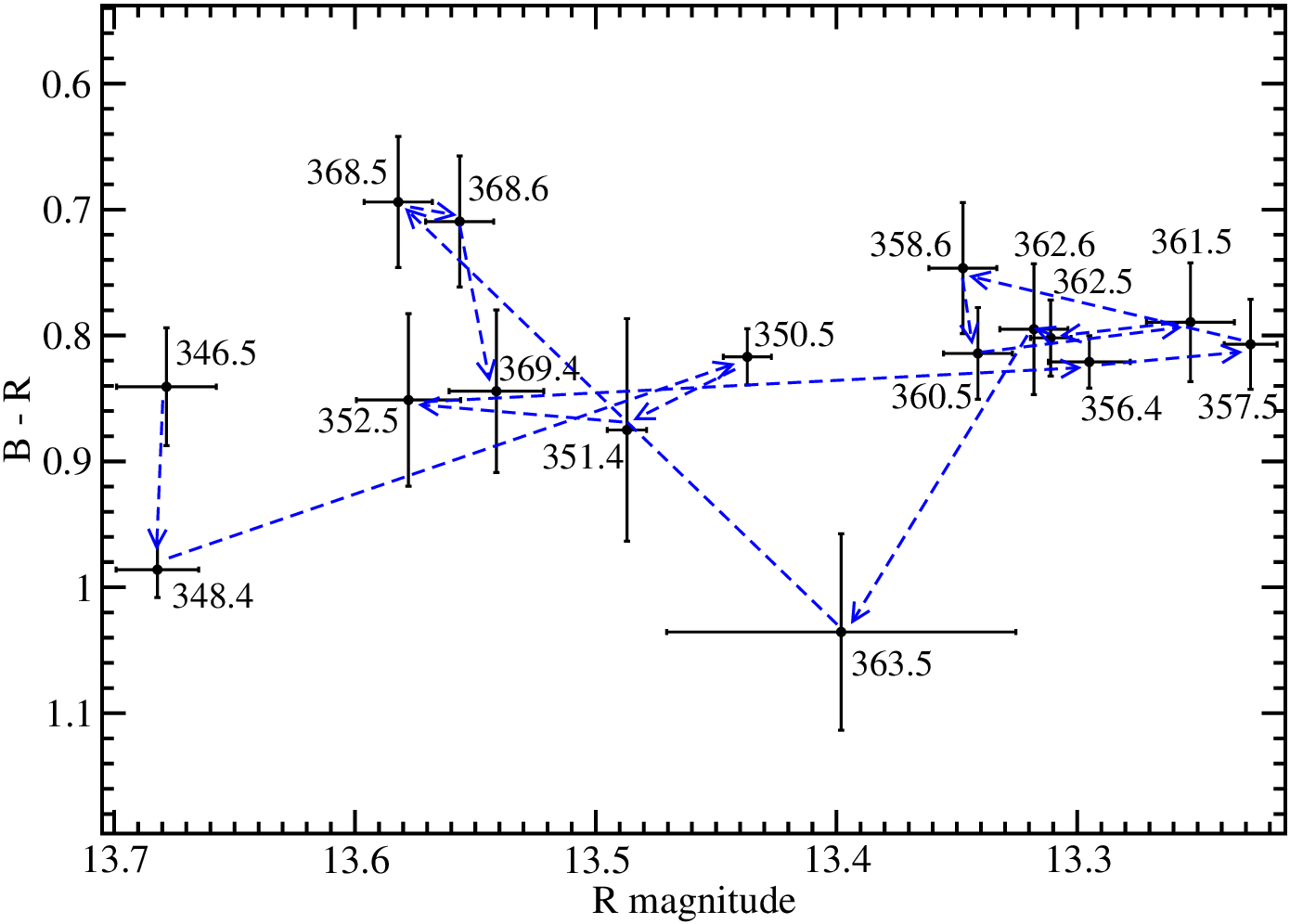}
\caption{Intrinsic optical hardness -- intensity diagram for the flare around 
Sept. 15, 2007. Individual points are labeled by JD - 2454000, and the arrows
indicate the time sequence. No hints of spectral hysteresis or other systematic 
spectral evolution are visible. }
\label{hysteresis}
\end{figure}

In \cite{boettcher05}, we had also found hints of systematic spectral 
evolution during optical flares, with peaks of the B - R hardness occurring 
consistently a few days before peaks in the R-band light curve. Hardness
-- intensity diagrams of an individual flare showed a tentative hint of
spectral hysteresis. During our 2007-2008 campaign, we did not find such
indications. Fig. \ref{hysteresis} shows the time-resolved hardness-intensity
diagram and spectral evolution during the exceptionally bright outburst
around Sept. 15, 2007 ($\sim$~JD~2454348 -- 2454370). The points are
labeled by their respective JD, and the arrows indicate the time sequence.
Our results do not confirm the existence of spectral hysteresis in 3C~66A.

\begin{figure}
\plotone{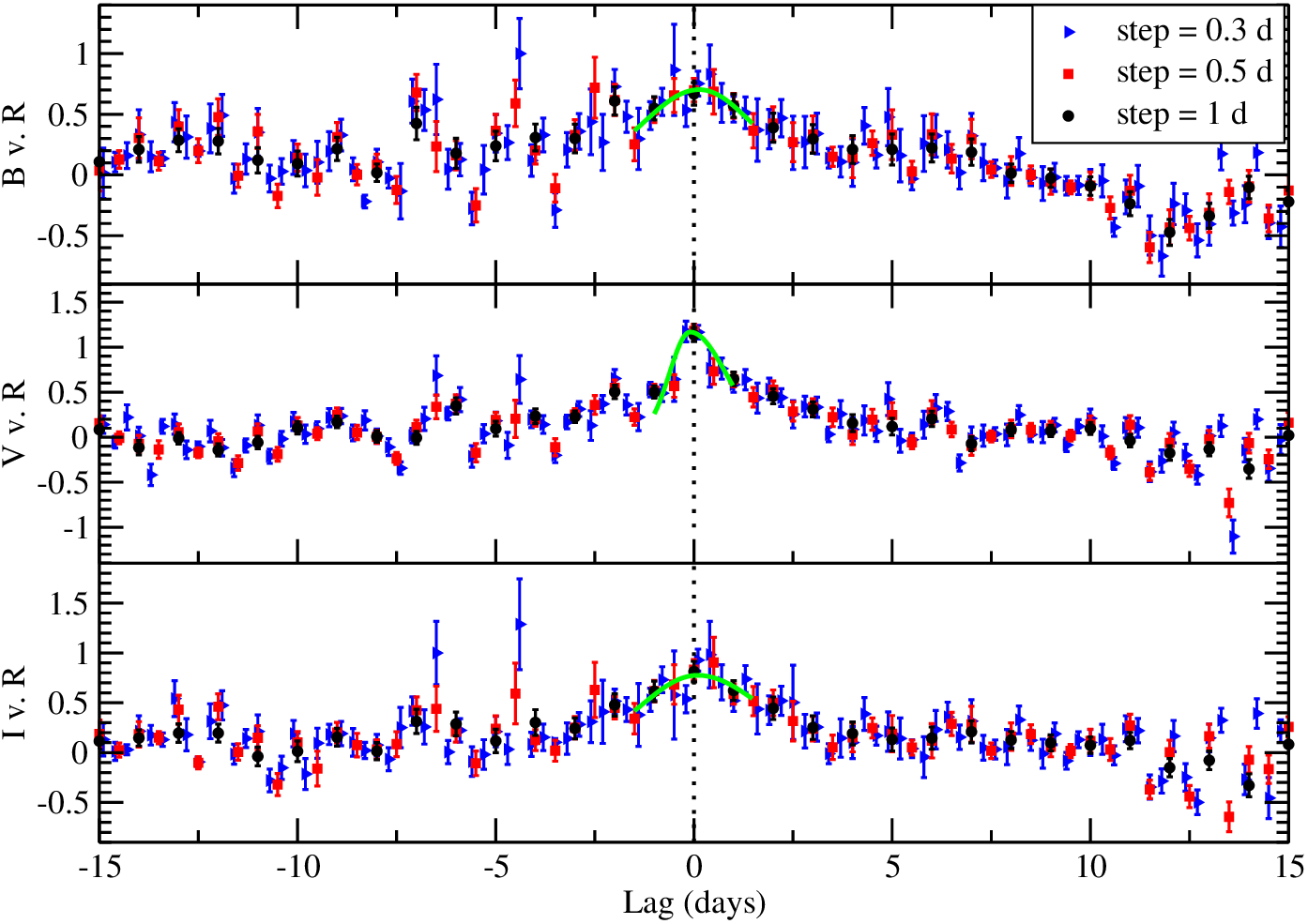}
\caption{Discrete correlation functions (DCF) of B, V, and I band vs. 
R band light curves. Different symbols indicate different time lag bin
sizes, as indicated in the legend, indicating that our results are independent
of our choice of a bin size. The curves indicate our best fit with an asymmetric
Gaussian.}
\label{DCF}
\end{figure}

\section{\label{crosscorrelations}Inter-band cross-correlations and time lags}

Our spectral variability analysis in the previous section already indicated
the lack of systematic spectral variability. This would be consistent with
a tight correlation at zero time lag between different optical bands. We
tested this hypothesis using the discrete correlation function \citep[DCF,][]{ek88}.

\begin{deluxetable}{ccccc}
\tabletypesize{\scriptsize}
\tablecaption{Best-fit parameters of an asymmetric Gaussian (see Eq. \ref{DCFfit}) 
to the discrete correlation functions between various optical bands. }
\tablewidth{0pt}
\tablehead{
\colhead{Band (vs. R)} & \colhead{B} & \colhead{V} & \colhead{I} 
}
\startdata
$F_0$&			0.703&		1.168&		0.776\\
$\sigma_r$ [d]&		$2.01 \pm 0.51$& $0.74 \pm 0.08$& $2.02 \pm 0.55$\\
$\sigma_d$ [d]&		$1.88 \pm 0.20$& $1.28 \pm 0.06$& $2.30 \pm 0.27$\\
$\tau_p$ [d]&         	$0.13 \pm 0.19$& $-0.01 \pm 0.05$& $0.09 \pm 0.21$\\
\enddata
\label{Gaussfits}
\end{deluxetable}

Fig. \ref{DCF} shows a compilation of the DCFs of the B, V, and I 
bands with respect to R as a reference band. We calculated DCFs with 
various time lag bins and obtained consistent results, independent of 
our choice of the bin size. In order to test for possible time lags 
between the various light curves, we fitted the DCFs with an asymmetric 
Gaussian of the form

\begin{equation}
DCF (\tau) = F_0 \, \cases{
e^{-{(\tau - \tau_p)^2 \over \sigma_{r}^2}} & for $\tau \le \tau_p$ \cr
e^{-{(\tau - \tau_p)^2 \over \sigma_{d}^2}} & for $\tau > \tau_p$\cr}
\label{DCFfit}
\end{equation}
where $F_0$ is the peak value of the DCF, $\tau_p$ is the delay
time scale at which the DCF peaks, and $\sigma_{r,d}$ parametrize the
Gaussian width of the DCF.
The best-fit functions are plotted in Fig. \ref{DCF},
and the best-fit parameters are quoted in Tab. \ref{Gaussfits}. These
results confirm that there is no evidence for time lags among the
optical bands in our campaign data. The rather low normalization of
the resulting DCF fits of $\sim 0.7$ and $\sim 0.8$ in the B vs. R
and I vs. R correlations might be a consequence of uneven sampling
with several substantial data gaps in the B and I light curves
(see Fig. \ref{timeline}).

\section{\label{discussion}Discussion}

The location of the synchrotron peak in the optical regime allows an estimate
of a combination of the magnetic field $B \equiv B_G$~G and the Doppler factor
$D \equiv 10 \, D_1$. We follow the approach of Eq. (8) in \cite{boettcher03}.
Here, the magnetic field energy density is parametrized as a fraction
$e_B$ of the energy density in relativistic leptons. Furthermore, constraining
the size of the emission region by the observed variability time scale,
$\tau_{\rm var, obs} \equiv 1 \, \tau_h$~hr, through $R = c \, \tau_{\rm
var, obs} \, D / (1 + z)$, we find

\begin{equation}
B_{e_B} = 8.4 \, \left( {d_{27}^4 \, f_{-10} \, e_B^2 \, (1 + z)^2 \over
\epsilon_{\rm sy, -6} \, (p - 2) \, \tau_h^6 \, D_1^{13}} \right)^{1/7}
\; {\rm G}
\label{BeB}
\end{equation}
where $d_{27} = 7.38$ is the luminosity distance in units of $10^{27}$~cm,
$f_{-10} \approx 0.5$ is the peak $\nu F_{\nu}$ flux in units of $10^{-10}$
ergs~cm$^{-2}$~s$^{-1}$, and $\epsilon_{\rm sy, -6}
= \epsilon_{\rm sy}/10^{-6} \approx 5$ the dimensionless photon energy,
$\epsilon = h \nu / (m_e c^2)$, at the peak of the synchrotron component
at $\nu_{\rm sy} \sim 5 \times 10^{14}$~Hz.
$p$ is the spectral index of the nonthermal electron population radiating
above the synchrotron peak. With these values, Eq. (\ref{BeB}) yields

\begin{equation} 
B_{e_B} = 19 \, e_B^{2/7} \, (p - 2)^{-1/7} \, \tau_h^{-6/7} \, D_1^{-13/7}
\; {\rm G}.
\label{BeBest}
\end{equation}

The lack of spectral variability may be interpreted as evidence that the
synchrotron cooling time of electrons radiating in the optical regime
is shorter than the light crossing time, which (including a factor $D/(1+z)$)
then determines the shortest observed variability time scale. If this were
not the case, frequency-dependent radiative cooling effects would be expected 
to lead to observable time lags and systematic spectral variability. Hence,
the condition ${\tau'}_{\rm sy} \, (1 + z) / D \le \tau_{\rm var, obs}$
may be translated into a Doppler-factor dependent lower limit on the magnetic 
field. Regarding electrons emitting in the R band, we find

\begin{equation}
\tau_{\rm sy, obs} = 7.32 \times 10^4 \, \left( {1 + z \over D} \right)^{1/2} 
\, B_G^{-3/2} \; {\rm s} \le \tau_{\rm var, obs}
\label{tausy}
\end{equation}
which yields the magnetic-field limit

\begin{equation}
B \ge 3.9 \, D^{-1/3} \, \tau_h^{-2/3} \; {\rm G}.
\label{Btau}
\end{equation}
Combining this limit with the magnetic-field estimate of Eq. (\ref{BeBest}),
we infer an upper limit on the Doppler factor:

\begin{equation}
D \le 28 \, (p - 2)^{-3/32} \, \tau_h^{-1/8}\, e_B^{3/16}.
\label{Dlimit}
\end{equation}
This restricts the Doppler factor to values typically found in the modeling
of blazar-type quasars and low-frequency peaked BL~Lac objects and excludes
extreme values of $\Gamma \sim D \gs 50$, found in one-zone, leptonic
synchrotron-self-Compton models for several high-frequency peaked BL~Lac 
objects detected at $E \gs 1$~TeV \citep[e.g.,][]{bfr07,gt08,finke08b}. 
This aspect is particularly exciting now, after the recent detection
of 3C~66A at $> 100$~GeV energies by VERITAS \citep{swordy08}. The very
high Doppler factors mentioned above are primarily required to accomodate
the very rapid variability, on time scales down to just a few minutes, at
GeV -- TeV energies seen in a few HBLs. Should future VHE $\gamma$-ray
observations of 3C~66A indicate such rapid variability for this object
as well, this would pose serious problems for conventional blazar emission 
models.

\section{\label{summary}Summary}

We reported on an extensive optical, near-IR (JHK), and radio monitoring 
campaign by the WEBT throughout the fall and winter of 2007 -- 2008,
prompted by a high optical state in September 2007 with $R < 13.5$. 
Twenty-eight observatories in 12 countries in North America, Europe,
and Asia contributed to this campaign.
 
The source remained in a high optical state throughout the observing
period and exhibited several bright flares on time scales of $\sim 10$~days, 
including an exceptional outburst around September 15 -- 20, 2007, reaching
a peak brightness at $R \sim 13.4$. Our campaign revealed microvariability 
with flux changes up to $\vert {\rm d}R/{\rm d}t  \vert \sim 0.02$~mag/hr with occasional
indications of a slower rise than decay of individual flares. 

Our observations do not reveal evidence for systematic spectral variability
in the overall high state ($R_{\rm dered} \ls 14.0$) covered by our 
campaign, in agreement with previous results for the same overall brightness 
state. Note, however, that a positive correlation between optical spectral 
hardness (B - R color index) and optical flux at dereddened R-band magnitudes 
of $R_{\rm dered} \gs 14.0$ has been found before. In particular, we do 
not find evidence for spectral hysteresis in 3C~66A for which hints were found 
in an earlier campaign in a somewhat lower flux state. On the same note, we
did not find evidence for time lags between optical bands. 

During our campaign, as in earlier campaigns, the peak of the low-frequency
(presumably synchrotron) component of the SED of 3C~66A was located in the
optical regime. From the observed synchrotron peak flux, we infer a value 
of the magnetic field of the emission region of $B \sim 19 \, e_B^{2/7} \, 
\tau_h^{-6/7} \, D_1^{13/7}$~G. From the lack of systematic spectral variability, 
a Doppler-factor dependent lower limit on the magnetic field can be derived, 
which implies an upper limit on the Doppler factor, $D \le 28 \, \tau_h^{-1/8} 
\, e_B^{3/16}$. This excludes extreme values of the Doppler factor of $D \gs 50$,
inferred for some high-frequency-peaked BL Lac objects detected at TeV energies.

\acknowledgments
The work of M. B\"ottcher was partially supported through NASA's XMM-Newton guest 
observer program, award no. NNX08AD67G.
The UMRAO is partially supported by funds from the NSF, most recently AST-0607523,
and from the University of Michigan's Department of Astronomy.
This publication is partly based on observations with the Medicina and Noto 
telescopes operated by INAF - Istituto di Radioastronomia.
The Mets\"ahovi radio astronomy team acknowledges support from the Academy
of Finland.
J. Wu and X. Zhou are supported by the Chinese National Natural Science 
Foundation grants 10603006, 10573020, and 10633020.
The AZT-24 observations were made within an agreement between the Pulkovo, 
Rome, and Teramo Observatories.

\end{document}